\documentclass[12pt]{article}
\usepackage{hyperref,amssymb,amsmath,graphicx,verbatim,eufrak,amsthm}
\usepackage{fullpage}
\usepackage{color}
\newtheorem{thm}{Theorem}%[section]
\newtheorem{cor}[thm]{Corollary}
\newtheorem{lem}[thm]{Lemma}

\newtheorem{clm}[thm]{Claim}

\newtheorem*{thm*}{Theorem}

\theoremstyle{definition}
\newtheorem{dfn}[thm]{Definition}
\theoremstyle{remark}

\numberwithin{equation}{section}

\newcommand{\br}[1]{\left[#1\right]}

\newcommand{\sr}[1]{\left(#1\right)}

\newcommand{\N}{N}
\newcommand{\B}{U}

\newcommand{\R}{\mathbb{R}}
\newcommand{\C}{\mathcal{C}}

\newcommand{\eps}{\varepsilon}

\DeclareMathOperator{\E}{\mathbb{E}}     % Without under-subscripts
\DeclareMathOperator{\Var}{Var}
\renewcommand{\Pr}{}
\let\Pr\relax
\DeclareMathOperator{\Pr}{\mathbb{P}}

\def\squareforqed{\hbox{\rlap{$\sqcap$}$\sqcup$}}
\def\qed{\ifmmode\squareforqed\else{\unskip\nobreak\hfil
\penalty50\hskip1em\null\nobreak\hfil\squareforqed
\parfillskip=0pt\finalhyphendemerits=0\endgraf}\fi}

\newcommand{\F}{\mathbb{F}}

\newcommand{\var}{\mathrm{var}}
\newcommand{\depth}{\mathrm{depth}}
\newcommand{\Inf}{\mathrm{Inf}_{\infty}}
\newcommand{\Infi}{\mathrm{Inf}}

\newcommand{\ignore}[1]{}

\newcommand{\sgn}{\mathrm{sgn}}
\newcommand{\MOD}{\mathrm{MOD}}

\begin{document}
\title{Polynomial Threshold Functions: Structure, Approximation and Pseudorandomness}
\author{
Ido Ben-Eliezer%
\thanks{School of Computer Science,
Raymond and Beverly Sackler Faculty of Exact Sciences, Tel Aviv
University, Tel Aviv, Israel. Email: idobene@tau.ac.il} \and
Shachar Lovett%
\thanks{Weizmann Institute of Science, Rehovot, Israel.
Email: shachar.lovett@weizmann.ac.il. Research supported by the Israel Science Foundation (grant 1300/05)}
\and
Ariel Yadin%
\thanks{Centre for Mathematical Sciences,
Wilberforce Road, Cambridge CB3 0WB, UK.
Email: a.yadin@statslab.cam.ac.uk}}

\maketitle

\begin{abstract}
We study the computational power of polynomial threshold functions,
that is, threshold functions of real polynomials over the boolean
cube. We provide two new results bounding the computational power of this model.

Our first result shows that low-degree polynomial threshold functions cannot
approximate any function with many influential variables. We provide a couple
of examples where this technique yields tight approximation bounds.

Our second result relates to constructing pseudorandom generators fooling
low-degree polynomial threshold functions. This problem has received attention
recently, where Diakonikolas {\em et al}~\cite{DGJSV09} proved that $k$-wise
independence suffices to fool linear threshold functions.
We prove that any low-degree polynomial threshold function, which can be represented
as a function of a small number of linear threshold functions, can also be fooled
by $k$-wise independence. We view this as an important step towards fooling
general polynomial threshold functions, and we discuss a plausible approach
achieving this goal based on our techniques.

Our results combine tools from real approximation theory,
hyper-contractive inequalities and probabilistic methods. In particular, we develop
several new tools in approximation theory which may be of independent interest.
\end{abstract}

\newpage

\section{Introduction}

A boolean function $h:\{-1,1\}^n \to \{-1,1\}$ is a threshold (or
sign) function of a real function $f:\{-1,1\}^{n} \to \R$ if
$$h(x_1,\ldots,x_n) = \sgn(f(x_1,\ldots,x_n)).$$
In this work we study thresholds of low-degree polynomials, or Polynomial Threshold Functions (PTFs). There is a long line of research that study the case of linear functions, i.e. degree~$1$ polynomials, which
are commonly called Linear Threshold Functions (LTFs), or {\em halfspaces}
(see, e.g., ~\cite{HMPST93,BGP07,DGJSV09} and their references within). A key
example for an LTF is the {\em majority} function which can be
defined as
$$
\mathrm Maj(x_1,\ldots,x_n) = \sgn(x_1+\ldots+x_n-\lceil n/2 \rceil).
$$

The main challenge that we tackle in our work is bounding the computational power
of low-degree PTFs. We consider two main problems. Constructing
explicit pseudorandom distributions that fool low-degree PTFs, and
providing lower bounds for the computation and approximation capabilities of PTFs.

\paragraph{Pseudorandom generators for PTFs}
An important question is whether $k$-wise independence fools PTFs for small values of $k$. In particular it is
interesting whether $k$ can be independent of the number of
variables $n$.

A boolean function $h:\{-1,1\}^n \to \{-1,1\}$ is $\eps$-fooled by
$k$-wise independence if for any $k$-wise independent distribution
$K$ taking values in $\{-1,1\}^n$ we have
$$
|\Pr_{x \in K}[h(x)=1] - \Pr_{x \in U}[h(x)=1]| \le \eps,
$$
where $U$ denotes the uniform distribution over $\{-1,1\}^n$. We say that a $k$-wise independence fools degree-$d$ polynomials if it fools any
threshold function $h(x)=\sgn(f(x)-t)$ for $t \in \R$), for any degree-$d$ real polynomial. This
notion can be extended to fooling real functions.
% Ido : This is not needed here, and can be found in the preliminaries section
%
% We say $k$-wise
%independence $\eps$-fools a real function $f:\{-1,1\}^n \to \R$ if
%for any $k$-wise disribtution $K$ and any $t \in \R$,
%$$
%|\Pr_{x \in K}[f(x) \ge t] - \Pr_{x \in U}[f(x) \ge t] | \le \eps.
%$$
%This is equivalent to $k$-wise independence $\eps$-fooling all
%threshold functions $h(x) = \sgn(f(x)-t)$ for all $t \in \R$.

The problem of whether $k$-wise independence fools LTFs was first
addressed by Benjamini et al.~\cite{BGP07}, who proved that
$k$-wise independence fools the majority function, and
subsequently by Diakonikolas et al.~\cite{DGJSV09} who proved
that $k$-wise independence fools LTFs. In both cases
$k=\mathrm{polylog}(\eps) \cdot \eps^{-2}$ was required to achieve
error $\eps$.

Our first result extends the result of Diakonikolas et al.~\cite{DGJSV09} to thresholds of low-degree polynomials which
depend on a small number of linear functions. We see it as an
important step towards building pseudorandom generators fooling
general PTFs. For a real polynomial $p(x) = \sum p_I \prod_{i \in
I} x_i$ define its {\em weight} as the sum of the absolute values of the
coefficients, excluding the constant coefficient, that is
$$
wt(p) = \sum_{I \ne \emptyset} |p_I|
$$

%\ignore{
%\begin{dfn}[$\delta$-normal functions]
%A function $f:\{-1,1\}^n \to \R$ is $\eps$-normal if the cdf of
%the distribution of $f$ over a uniform input is $\eps$-close to
%the normal distribution, that is, for any $t \in R$
%$$
%|\Pr_{x \in \{-1,1\}^n}[f(x) \ge t] - \Pr[N \ge t]| \le \eps
%$$
%where $N~N(0,1)$ is a standard normal variable.
%\end{dfn}
%|}

\begin{thm}\label{thm:intro:prg_poly_few_linear}
Let $f:\{-1,1\}^n \to \R$ be a degree~$d$ polynomial, which can be
decomposed as a function of $m$ linear functions. That is, there
exist linear functions $g_1,\ldots,g_m:\{-1,1\}^n \to \R$ and a
degree-$d$ polynomial $p:\R^m \to \R$ such that
$$
f(x) = p(g_1(x),\ldots,g_m(x))
$$
for all $x \in \{-1,1\}^n$. Assume that $g_1,\ldots,g_m$ are normalized
such that $\E[g_1^2]=\ldots=\E[g_t^2]=1$. Then $k$-wise
independence $\eps$-fools $f(x)$ for $$k =
\mathrm exp(O(d/\eps)^d) + \mathrm poly((\log{m} \cdot d/\eps)^d, m, wt(p)).$$
\end{thm}

%\ignore{
%A special case is a function which depends on a few LTFs, for
%example intersection of halfspaces. In this special case we show
%that the techniques of Diakonikolas et al.~\cite{DGJSV09} can be
%used to show that $k$-wise independence fools such functions.
%
%\begin{thm}\label{thm:intro:prf_func_few_LTF}
%Let $h_1,\ldots,h_m:\{-1,1\}^n \to \{-1,1\}$ be LTFs. Let
%$F:\{-1,1\}^n \to \{-1,1\}$ be some function of $h_1,\ldots,h_m$,
%i.e.
%$$
%F(x) = H(h_1(x),\ldots,h_m(x))
%$$
%where $H:\{-1,1\}^m \to \{-1,1\}$ is arbitrary. Then $k$-wise
%independence fools $F$ for $k=XXX$.
%\end{thm}
%}

\paragraph{Lower bounds for approximation by PTFs}

A boolean function $g:\{-1,1\}^n \to \{-1,1\}$ is said to be
$\eps$-approximated by degree~$d$ PTFs, if there exists a
degree~$d$ PTF $h(x)$ s.t. $\Pr_{x \in U}[h(x)=g(x)] \ge
1-\eps$.

We prove that functions whose variables have high influence cannot
be approximated by low-degree PTFs, where the influence of a variable
$x_i$ in $g$ is defined as the probability that flipping $x_i$ changes
the value of $g$, i.e.
$$
\Infi_i(g) = \Pr_x[g(x) \ne g(x \oplus e_i)],
$$
where $e_i$ is the $i$-th unit vector. We prove

\begin{thm}\label{thm:intro:ptf_no_approx}
Let $g:\{-1,1\}^n \to \{-1,1\}$ be a boolean function, such that
$\Infi_i(g) \ge \tau$ for at least $n^{\alpha}$ variables. Then for any degree-$d$
polynomial threshold function $h$ we have
$$
\Pr_{x}[h(x)=g(x)] \le 1 - \frac{\tau}{2} + \eta
$$
where $\eta = O(d / (\alpha \log{n})^{1/8d})$.
\end{thm}

We illustrate the power of Theorem~\ref{thm:intro:ptf_no_approx} by
showing two examples. The first one shows that $\MOD_m$ function cannot be approximated by low degree PTFs, while the second result shows that any low-degree polynomials over
$\F_2$ cannot be approximated by low-degree PTFs much better than
the best trivial approximation. Let define the $\MOD_m$ function as
$$
\MOD_m(x_1,\ldots,x_n) = \bigg\{
\begin{array}{cc}
  1 &  \sum_{i=1}^{n} \frac{x_i+1}{2}  \equiv 0 \pmod m\\
  -1 & \sum_{i=1}^{n} \frac{x_i+1}{2}   \not\equiv 0 \pmod m\\
\end{array}
$$
Note that as $\frac{x_i+1}{2} \in \{0,1\}$, this definition is essentially equivalent to the common one. We have the following.

\begin{cor}\label{cor:intro:modm_no_approx}
Let $h:\{-1,1\}^n \to \{-1,1\}$ be a degree-$d$ polynomial
threshold function for \\
$d \le O(\log\log{n} / \log\log\log{n})$.
Then
$$
\Pr[h(x) = \MOD_m(x)] \le 1 - \frac{1}{m} + o(1).
$$
\end{cor}

This result is tight in the sense that trivially the $MOD_m$ function admits an $1-\frac{1}{m}$ approximation by the constant $-1$ function (which is also a degree-$0$ PTF).

\begin{cor}\label{cor:intro:poly_no_approx}
Let $q:\{-1,1\}^n \to \{-1,1\}$ be a degree-$r$ polynomial over $\F_2$ depending on all variables.
Let $h:\{-1,1\}^n \to \{-1,1\}$ be a degree-$d$ polynomial
threshold function for $d \le O(\log\log n / \log\log\log n)$. Then
$$
\Pr[h(x) = q(x)] \le 1 - 2^{-r} + o(1).
$$
\end{cor}
This result is essentially tight, as if $q$ is a product of $r$ linear forms, then
the constant $1$ function gives an $1-2^{-r}$ approximation of $q$.

\subsection{Tools}

\paragraph{Approximation tools and $k$-wise independence.}
Several recent works used the method of approximating by real polynomials to show that certain families of functions are fooled by $k$-wise independent distributions. This method can be described as follows. In order to show that $k$-wise independence $\eps$-fools a certain family of functions, one has to show that for every function $f$ in that family, there is a degree $k$ polynomial $p_l$ and degree $k$ polynomial $p_u$, such that for every $x \in \{-1,1\}^n$ we have $ p_l(x) \leq f(x) \leq p_u(x)$, and such that $\E_x[p_u(x)-p_l(x)] \leq \eps$. Using this technique, Bazzi~\cite{Bazzi09} proved in a breakthrough paper that logarithmic-wise independence fools DNF and CNF formulas. Later, Braverman~\cite{Braverman09} proved that polylogarithmic-wise independence fools small constant depth circuits, settling a conjecture of Linial and Nisan~\cite{LN90}.

In this work we use the method of approximating polynomials for the problem of fooling low degree PTFs. We introduce a general method of obtaining polynomials which are both bounding and approximating for any function
which depends on a small number of subfunctions whose tail distribution `behaves nicely'. In our case we apply it for functions of a few linear functions, but we believe that these methods should have independent interest.

Our starting point is the multidimensional Jackson's theorem, which states that every Lipschitz function $f$ on $m$ variables admits an $\eps$-approximation by a degree-$d$ polynomial, where $d$ depends only on $\eps$, $m$ and the Lipschitz constant of $f$. We then use several additional techniques to show that $f$ admits a polynomial approximation $p$ which is a good approximation in a multidimensional box near the origin, and above $f$ everywhere. Finally, we apply these techniques as well as some concentration and anti-concentration results to show that $p$ is a good approximation for $f$.

Finally, we apply these techniques to show that any threshold of a function of a few linear functions (or a function of a few linear PTF's) can be fooled by $k$-wise independence, for $k$ that is independent of the number of variables.

\paragraph{Decision trees and approximation of PTF.} Our first tool is a new structural result about PTFs. Given a polynomial threshold function $p$, we show that it has a small set of variables, on which {\em most} of their possible assignments we obtain a function with no influential variable. More precisely, the partial assignments are given by a small depth decision tree.

Let $D$ be a decision tree on the variables $x_1,\ldots,x_n$. Each internal node of $D$ is labeled by some variable and has two outgoing edges, corresponding to
the possible assignments to this variable. The set of leaves of the decision tree correspond to partial assignments
to the variables. The set of the leaves of $D$ is denoted by $L(D)$, and for any $\ell \in L(D)$ and a function
$f(x_1,\ldots,x_n)$ we denote by $f|_{\ell}$ the function restricted to the partial assignment given by $\ell$. For more precise definitions see Section~\ref{sec:prelim}. We prove the following result.

\begin{lem}\label{lem:small_inf_by_decision_tree}
Let $f:\{-1,1\}^n \to \R$ be a degree-$d$ polynomial, and let
$h(x) = \sgn(f(x))$. For any $\epsilon,\delta>0$, there exists a
decision tree $D$ of depth at most $2^{e d/\delta} \cdot
\log(1/\epsilon)$, such that
$$
\Pr_{\ell \in L(D)}[\Inf(f|_{\ell}) > \delta] < \epsilon
$$
and
$$
\Pr_{\ell \in L(D)}[\Inf(h|_{\ell}) > \delta'] < \epsilon
$$
for $\delta' = O(d \cdot \delta^{1/8d})$.
\end{lem}

We sketch the proof of Theorem~\ref{thm:intro:ptf_no_approx}.
If a function $g$ approximates a PTF $h$, then after most partial assignments of variables, $g$ still approximates $h$. We show that under most of these assignments, our obtained PTF does not have any influential variable, and therefore cannot approximate functions with many influential variables.

Independently of our work, Diakonikolas et al.~\cite{DSTW09} and Harsha et al.~\cite{HKM09}
proved similar results. We state their results in our terminology.

\begin{thm}[Theorem 1 in~\cite{DSTW09}]\label{thm:DSTW_regularity}
Let $f:\{-1,1\}^n \to \R$ be a degree-$d$ polynomial, and let
$h(x) = \sgn(f(x))$. For any $\tau > 0$, there exists a decision tree $D$ of depth
$\frac{1}{\tau} \cdot (d \log{\tfrac{1}{\tau}})^{O(d)}$ such that
with probability $1-\tau$ over a random leaf $\ell \in L(D)$, the function $h|_{\ell}$
is either $\tau$-close to being constant, or has $\Inf(h) < \tau$.
\end{thm}

\begin{thm}[Lemmas 5.1 and 5.2 in~\cite{HKM09}]\label{thm:HKM_regulariy}
Let $f:\{-1,1\}^n \to \R$ be a degree-$d$ polynomial, and let
$h(x) = \sgn(f(x))$. For any $\tau > 0$, there exists a decision tree $D$ of depth
$\frac{\mathrm{polylog(\tau)}}{\tau^2} \cdot exp(d)$ such that
with probability $1-\tau$ over a random leaf $\ell \in L(D)$, the function $h|_{\ell}$
is either $\tau$-close to being constant, or has $\Inf(h) < \tau$.
\end{thm}

We note that using Theorem~\ref{thm:HKM_regulariy} instead of Lemma~\ref{lem:small_inf_by_decision_tree}
one can get an improvement in the dependence on the degree in Theorem~\ref{thm:intro:ptf_no_approx}.
In particular, Corollaries~\ref{cor:intro:modm_no_approx} and~\ref{cor:intro:poly_no_approx} hold
for degrees $d \le O(\log{n} / \log\log{n})$.

\subsection{Towards fooling low degree PTFs}
We propose a general method for proving that $k$-wise independence fools low degree PTFs.
This is a high level approach and currently we are able to prove only a special case.

Let $f:\{-1,1\}^n \to \R$ be a real function. We say that $f$ is $\delta$-normal if the distribution of $f(x)$
over uniform input is $\delta$-close to the standard normal distribution. That is,
$$
|\Pr_{x \in U}[f(x) \ge t] - \Pr[N \ge t]| < \delta
$$
for any $t \in \R$, where $N \sim N(0,1)$ is a standard normal variable. In what follows we let
$f(x)$ be a degree~$d$ polynomial, $h(x)=\sgn(f(x))$ a PTF and $\eps>0$ the required error.

\begin{enumerate}
    \item \textbf{Reduction to low-influence PTF}: It is enough to prove that $k$-wise independence
    fools PTFs with small influences. We prove this in Lemma~\ref{lem:small_inf_by_decision_tree} and
    Claim~\ref{clm:enough_fool_leaves}. The important properties of PTFs with low influences is that
    their distribution is not concentrated around any specific value (see Lemma~\ref{lem:poly_small_inf_anti_concentration}), which can later be used to build approximating
    polynomials for such functions.

    \item \textbf{$\delta$-normal polynomials}: Assume that $f(x)$ is a degree-$d$ polynomial with low influences which is $\delta(\eps)$-regular. Then $h(x)=\sgn(f(x))$ is fooled by $k(\eps)$-wise independence. This can be proved using the same proof technique of Diakonikolas et al.~\cite{DGJSV09}, using the approximating polynomials for the $\sgn$ functions they construct, when replacing the
        tail bounds for linear polynomials by the normal distribution.

    \item \textbf{Functions of a few $\delta$-normal polynomials}: Assume that $f(x)$ is a degree-$d$ polynomial with low influences, which can be decomposed as a function of $m$ polynomials $g_1,\ldots,g_m$, each is $\delta(m,\eps)$-normal. Then $h(x)=\sgn(f(x))$ is fooled by $k(m, \eps)$-wise independence. Our proofs can
        be slightly altered to prove this, again replacing
        tail bounds for linear polynomials by the normal distribution. This can be also extended when allowing
        a small error term.

    \item \textbf{Regularization of degree-$d$ polynomials}: We conjecture that for every $\delta,\tau>0$,
    any degree~$d$ polynomial $f:\{-1,1\}^n \to \R$ can be regularized in the following way. There exist a small number $t=t(d,\delta,\tau)$ of variables  $x_{i_1},\ldots,x_{i_t}$, and a small number
    $m=m(d,\delta,\tau)$ of $\delta$-normal polynomials $g_1,\ldots,g_m:\{-1,1\}^n \to \R$,
    a low-degree polynomial $p:\R^m \to \R$ and an error polynomial $e:\{-1,1\}^n \to \R$ with $\|e\|_2 < \tau$, such that
    $$
    f(x) = p(x_{i_1},\ldots,x_{i_t},g_1(x),\ldots,g_m(x)) + e(x).
    $$
    For linear polynomials, this can be proved using the tools of Diakonikolas et al.~\cite{DGJSV09}. We were able to prove this conjecture also for quadratic polynomials, and conjecture that the same holds for all constant degrees~$d$.

    \item \textbf{Putting everything together}: Let $f(x)$ be a degree~$d$ PTF. We start by reducing it to a PTF with
    low influences using a partial assignment for a small number of variables. We use the conjecture to decompose
    it as a function of a small number of $\delta$-normal PTFs, and use this decomposition to prove that $k$-wise independence to fool $f$.

\end{enumerate}

\noindent
\textbf{So where does this fail?} The critical point of failure is in the dependence of the number
of functions $m$ used in the decomposition of $f$, and the required distance $\delta$ between their
distribution and the normal distribution. We can prove that if $f$ can be decomposed into a function
of $m$ $\delta$-normal functions for small enough $\delta$ then the proof follows through. The problem is that
$\delta$ has to be very small; in particular $\delta < exp(-m^5)$. On the other hand in the regularization
conjecture, the number of components $m$ depend on $\delta$. We can prove the regularization conjecture
for quadratic polynomials for $m \ge 1/\delta^2$. These two requirements have no common solution.

We note the independently of our work, Meka and Zuckerman~\cite{MZ09} constructed an explicit pseudorandom
generator fooling all degree-$d$ PTFs. Their construction involves partitions the set of inputs into a small
number of buckets (using a pairwise independent hash function), and then applying $k$-wise independent distribution to each bucket independently.

\subsection{More related Work}

The study of distributions that fool low-degree polynomials and related functions
has received considerable attention. For example, fooling linear
polynomials over finite fields~\cite{NN90,AGHP92}, which has a
numerous number of applications and extensions, pseudorandom
generators for low degree
polynomials~\cite{BV07,Lovett08,Viola08,ABK08} and fooling modular
sums~\cite{LRTV09}.

Bruck~\cite{Bruck90} studied polynomial threshold functions, and proved that such functions can be computed by depth-$2$ polynomial sized circuits with unbounded fan-in linear threshold gates. Aspnes et al.~\cite{ABFR91} studied the approximation of boolean functions by some threshold functions. Namely, they study the best possible approximation for the parity function and other symmetric functions by low-degree PTF, and proved that for every degree-$k$ PTF $p$, we have
$$
\Pr_x[p(x) \neq \mathrm PARITY(x)] \geq \frac{\sum_{i=0}^{\lfloor (n-k-1)/2 \rfloor} {n \choose i}}{2^n},
$$
and this bound is tight. However, their bounds for other functions are not fully explicit and are not tight.

A few recent results consider the problem of constructing
pseudorandom generators for threshold functions. This problem has a
natural geometrical interpretation. Rabani and
Shpilka~\cite{RS09} provided a construction of $\eps$-net for
halfspaces. Namely, a set of points $S$ for which for every
halfspace $h$ that satisfies $\eps \leq \Pr_{x \in
\{-1,1\}^{n}}[h(x) = 1] \leq 1-\eps$ there are two points
$s_1,s_2 \in S$ such that $h(s_1) = -1$ and $h(s_2) = 1$. The size
of their construction is polynomial in $n$ and $\frac{1}{\eps}$.
~\cite{DGJSV09} proved that any $k$-wise distribution fools
halfspaces, for $k$ that is polynomial in $\frac{1}{\eps}$. Their
dependence on $k$ is nearly optimal, as shown by Benjamini et al.~\cite{BGP07}.

A subsequent work of Diakonikolas et al.~\cite{DKN09} show that $k$-wise independence fools
quadratic threshold functions, and intersections of such functions.

The rest of our paper is organized as follows. We introduce some preliminary definitions and tools in Section~\ref{sec:prelim}. This section includes definitions and results that are related to $k$-wise independence, decision trees, concentration of multivariate polynomials and some other analytical tools.
In Section~\ref{sec:structure_results} we present our new structural results on low-degree PTF, and present our application that shows that certain functions cannot be approximated by low degree PTF. Finally, in Section~\ref{section-approx-and-fool} we present our new tools from approximation theory, and show that $k$-wise independence fools thresholds of functions of a few linear polynomials.

Throughout this work we do not try to optimize constants. Also, we omit floor and ceiling signs whenever these are not crucial.

\section{Preliminaries}
\label{sec:prelim}

In this section we provide some necessary definitions that will be
widely used throughout the work, including definitions and tools related to $k$-wise independent distributions, decision trees, analytical tools, and concentration bounds for multivariate polynomials.

\subsection{$k$-wise independent distributions and polynomials}
\label{subsection-pre-basic}

A distribution $D$ on the boolean cube $\{-1,1\}^{n}$ is $k$-wise
independent if the marginal distribution of any $k$ coordinates
is the uniform distribution. There are
explicit constructions of such distributions of size $O(n^{\lceil
k/2 \rceil})$, and these constructions are essentially
optimal~\cite{ABI86}.

Given a class of functions $\mathbb S$ from the boolean cube to
$\{-1,1\}$, a distribution $D$ $\eps$-fools $\mathbb S$ if for
every $\varphi \in \mathbb S$, we have $$|\Pr_{x \in
U}[\varphi(x)=1]-\Pr_{x \in D}[\varphi(x)=1]| \leq \eps.$$

Combining these two definitions, for simplicity we define the following.

\begin{dfn}[$k$-wise independence fooling boolean functions]
A boolean function $f:\{-1,1\}^n \to \{-1,1\}$ is said to be
fooled by $k$-wise independence with error $\eps$, if for any
$k$-wise independent distribution $K$,
$$
|\Pr_{x \in U}[f(x)=1] - \Pr_{x \in K}[f(x)=1]| \le \eps.
$$
\end{dfn}

The following claim is sufficient for $k$-wise distributions to
$\eps$-fool a boolean function.

\begin{clm}\label{clm:kwise_poly}
Let $f:\{-1,1\}^n \to \{-1,1\}$. Assume there are two degree-$k$
polynomials $p_u,p_l:\{-1,1\}^n \to \R$ such that
\begin{itemize}
    \item $p_l(x) \le f(x) \le p_u(x)$ for all $x \in \{-1,1\}^n$.
    \item $\E_{x \in \B}[p_u(x) - p_l(x)] \le \eps$.
\end{itemize}
Then $k$-wise independence fools $f$ with error $\eps$.
\end{clm}

The proof of this claim is simple, and can be found for example in~\cite{Bazzi09}. It is worth noting that Bazzi~\cite{Bazzi09} also proved that the condition is necessary using linear programming duality.

Our next definition extends the notion of fooling boolean functions, and defines it for real functions as well.

\begin{dfn}[$k$-wise independence fooling real functions]
Let $f:\{-1,1\}^n \to \R$ be a function. We say that $k$-wise
distributions fool $f$ with error $\eps$, if for any $k$-wise
distribution $K$ over $\{-1,1\}^n$, and any $t \in \R$,
$$
|\Pr_{x \in U}[f(x) \le t] - \Pr_{x \in K}[f(x) \le t] | \le
\epsilon
$$
\end{dfn}

A real function $f(x_1,\ldots,x_n)$ is a degree-$d$ polynomial if
it can be represented as
$$
f(x) = \sum_{k=0}^{d} \sum_{i_1 \le \ldots \le i_k \in [n]}
\alpha_{i_1,\ldots,i_k} x_{i_1} \ldots x_{i_k}.
$$
A polynomial is {\em multilinear} if each variable appears in
every monomial at most once. Equivalently, it can be represented
as
$$
f(x) = \sum_{k=0}^{d} \sum_{i_1 < \ldots < i_k \in [n]}
\alpha_{i_1,\ldots,i_k} x_{i_1} \ldots x_{i_k}.
$$

Each function $f:\{-1,1\}^n \to \R$ can be uniquely represented by
a multilinear polynomial. We will interchangeably regard $f$ both
as a boolean function and as a multilinear polynomial.

\subsection{Decision trees}
\label{subsection-pre-tree}

A {\em Decision Tree} over binary variables $x_1,\ldots,x_n$ is a
binary tree, where each internal node $v$ is labeled by one of the
variables $x_v$, such that the labels along any path from the root to a
leaf are distinct. Also, the two (directed) edges that leave each
node are labeled by $-1$ and $1$. Therefore, given a path $P$ from
the root to a leaf, for every variable $x$ that appears along the
path we can uniquely define a value $x_P \in \{-1,1\}$ to be the
label of the edge in $P$ that leaves the node labeled by $x$.

A path $P$ from the root to a leaf $\ell$ defines a partial
assignment $A_{\ell}$ by assigning every variable that appears on
$x$ by $x_P$. All the variables that do not appear on $P$ remain
unassigned.

We denote the set of variables labeling the vertices in the path to
$\ell$ by $\var(\ell)$. We denote the set of leaves of a decision
tree $D$ by $L(D)$.

The {\em depth} of a leaf is the length of the path from the root to
it, and the depth of a decision tree is the maximal depth of a leaf.

With a slight abuse of notation, we define a random leaf in a
decision tree to be the result of the following procedure. We start
at the root, and at each step we move to one of his children,
uniformly and independently of the other choices. When we arrive a
leaf $\ell$ we output it. Equivalently, we choose each leaf $\ell$
with probability $2^{-\depth(\ell)}$.

We now can define the restriction of a function with respect to a
certain leaf $\ell$ and with respect to a decision tree $D$.

\begin{dfn}
Let $f:\{-1,1\}^{n} \to \R$ be a function, $D$ be a decision tree
on $x_1,\ldots,x_n$ and $\ell$ be a leaf in $D$. We define the
restriction of $f$ to $\ell$, denoted by $f|_{\ell}$, to be the
function obtained by $f$ after assigning the variables
$x_1,\ldots,x_n$ according to $A_{\ell}$. Namely, the domain of
$f|_{\ell}$ is $\{-1,1\}^{[n] \setminus \var(\ell)}$, and the range
of $f|_{\ell}$ is $\R$.

Similarly, given a distribution $\mathcal D$, define its restriction
to $\ell$, $\mathcal D|_{\ell}$ to be the the distribution obtained
from $D$ conditioning on the partial assignment $A_{\ell}$.

We define a random function $f|_{D}$ by choosing a random leaf
$\ell$ of $D$ and restricting $f$ to $\ell$.
\end{dfn}
%
%We finish this subsection by defining a distribution which is the
%restriction of a function to a decision tree.

%\begin{dfn}[Restriction of a function to decision tree]
%Let $f:\{-1,1\}^{n} \to \R$ be a function, and let $D$ be a
%decision tree on $x_1,\ldots,x_n$. For a leaf $\ell$ of $D$, we
%define the restriction of $f$ to $\ell$, denoted as $f|_{\ell}$,
%to be $f$ restricted to the partial assignment corresponding to
%$\ell$, so $f|_{\ell}:\{-1,1\}^{[n] \setminus \var(\ell)} \to \R$.
%
%Similarly for a distribution $X$ over $\{-1,1\}^{n}$ we define
%its restriction to $\ell$, denoted by $X_{\ell}$, to be the
%distribution on the subspace determined by $\ell$.
%\end{dfn}

We will need the following easy claim.
\begin{clm}\label{clm:enough_fool_leaves}
Let $f:\{-1,1\}^n \to \R$ be a function, and $D$ a decision tree,
such that
$$
\Pr_{\ell \in L(D)}\br{\textrm{$k$-wise independent distributions
fool $f|_{\ell}$ with error $\eps$}} \ge 1-\delta.
$$
Then $(k+\mathrm{depth}(D))$-wise independent distributions fool $f$ with
error $\eps+\delta$.
\end{clm}

\begin{proof}
Let $K$ be some $k'$-wise independent distribution for
$k'=k+\mathrm{depth}(D)$. For any leaf $\ell \in L(D)$, the restriction
$K|_{\ell}$ of $K$ given by $\ell$ is $k$-wise independent.

Let $\ell \in L(D)$ be a random leaf of $D$. Say $\ell$ is {\em
good} if $k$-wise independent distributions fool $f|_{\ell}$ with
error $\eps$. By our assumption $\ell$ is good with probability at
least $1-\delta$.

Let $t \in \R$. For any good leaf we have
$$
|\Pr_{x \in U|_{\ell}}[f(x) \le t] - \Pr_{x \in K|_{\ell}}[f(x)
\le t]| < \eps.
$$

For any other leaf we can bound
$$
|\Pr_{x \in U|_{\ell}}[f(x) \le t] - \Pr_{x \in K|_{\ell}}[f(x)
\le t]| \le 1.
$$

Hence we get
$$
|\Pr_{x \in U}[f(x) \le t] - \Pr_{x \in K}[f(x) \le t]| \le
\E_{\ell \in L(D)} |\Pr_{x \in U|_{\ell}}[f(x) \le t] - \Pr_{x \in
K|_{\ell}}[f(x) \le t]| \le \eps + \delta.
$$
\end{proof}

We will also require a bound on the $L_2$ norm of linear functions, under a partial restriction
given by a decision tree.
\begin{lem}\label{lem:L2_linear_decision_tree}
Let $g:\{-1,1\}^n \to \R$ be a linear function with $\E[g^2]=1$.
Let $D$ be a decision tree. Then
$$
\Pr_{\ell \in L(D)}[\E[(g|_{\ell})^2] \ge t] \le 3e^{-t/8}.
$$
\end{lem}

\begin{proof}
We will need the following variant of the Azuma-Hoeffding inequality.
Let $X_1,\ldots,X_n$ be random variables, such that $X_i = c_i(X_1,\ldots,X_{i-1})$ or $X_i = -c_i(X_1,\ldots,X_{i-1})$, each with probability $1/2$, where $c_i:\{-1,1\}^{i-1} \to \R$ is
some deterministic function, such that a.s. $X_1^2 + \ldots + X_n^2 \le 1$. We will prove that
$$
\Pr[X_1 + \ldots + X_n \ge t] \le e^{-t^2/2}.
$$

First we show how we apply this inequality. Let $g(x) = a + \sum a_i x_i$ where $\sum a_i^2 + a=1$.
Let $\ell$ be a leaf of $D$. Notice that $g|_{\ell}(x) = (a + \sum_{i \in \var{\ell}} a_i x_i|_{\ell}) + \sum_{i \notin \var{\ell}} a_i x_i$. Hence, to bound the probability that $\E[(g|_{\ell})^2]$ is large, we need to bound the probability that $\sum_{i \in \var{\ell}} a_i x_i|_{\ell}$ is large. We will assume w.l.o.g that $t \ge 8$ since otherwise the required inequality holds immediately.

Define a sequence of random variables $X_1,X_2,\ldots$. Let $i_1$ be the index of the first variable
queried by $D$. Define $X_1 = \pm a_{i_1}$. Given the value of $x_{i_1}$, let $i_2$ be the index
of the second variable queried by $D$. Define $X_2 = \pm a_{i_2}$. Notice that in fact $X_2 = \pm c_2(X_1)$.
Let $i_3$ be the index of the third variable queried by $D$, and define $X_3 = \pm a_{i_3}$. Again,
$X_3 = \pm c_3(X_1,X_2)$, and we continue until we reach a leaf. If $X_d$ is a leaf of $D$, we define
the remaining variables $X_{d+1},\ldots,X_n$ to be $0$. Let $X = \sum X_i$. Notice that
$$
g|_{\ell}(x) = (a + X) + \sum_{i \notin \var(\ell)} a_i x_i.
$$
Since the conditions of the inequality hold for $X_1,\ldots,X_n$, we get that $\Pr[X \ge t] \le e^{-t^2/2}$.
We wish to bound the probability over $\ell \in L(D)$ that $\E[(g|_{\ell})^2] \ge t]$. If this event occurs,
then we must have $X \ge \sqrt{t}-1$. Since we assume $t \ge 8$ this gives $X \ge \sqrt{t}/2$, which gives
$$
\Pr_{\ell \in L(D)}[\E[g|_{\ell}^2] \ge t] \le \Pr[X \ge \sqrt{t}/2] \le e^{-t/8}.
$$

We now turn to prove the modification of the Azuma-Hoeffding inequality.
Set $\lambda>0$ to be determined later, and consider $E=\E[e^{\lambda(X_1+\ldots+X_n)}]$.
We can decompose $E = \prod_{i=1}^k \E[e^{\lambda X_i}|X_1,\ldots,X_{i-1}]$. We have
$$
\E[e^{\lambda X_i}|X_1,\ldots,X_{i-1}] = \frac{1}{2} e^{\lambda c_i(X_1,\ldots,X_{i-1})} + \frac{1}{2} e^{-\lambda c_i(X_1,\ldots,X_{i-1})}.
$$
Using the inequality $\frac{1}{2}(e^x + e^{-x}) \le e^{x^2/2}$ we get
$$
\E[e^{\lambda X_i}|X_1,\ldots,X_{i-1}] \le e^{\lambda^2 c_i(X_1,\ldots,X_{i-1})^2 / 2}.
$$
Hence
\begin{align*}
E \le \E_{X_1,\ldots,X_n} [e^{\lambda^2/2 \cdot (c_1^2 + c_2(X_1)^2 + \ldots + c_n(X_1,\ldots,X_{n-1})^2)}]
= \E_{X_1,\ldots,X_n} [e^{\lambda^2/2 \cdot (X_1^2+\ldots+X_n^2)}] \le e^{\lambda^2/2}.
\end{align*}
Thus we get
$$
\Pr[X_1 + \ldots + X_n \ge t] \le e^{\lambda^2/2 - \lambda t}.
$$
Setting $\lambda = t$ gives the required inequality.
\end{proof}

% Ido : This not relevant anymore, defined in the introduction.
%\subsection{Influence of a function}
%\label{subsection-pre-inf}
%
%We next define the influence of a variable and of a function. The
%definition of the influence of a single variable extends the
%standard definition that is used in the context of boolean
%functions. We say that a function has high max influence if it has a
%variable with high influence.
%
%\begin{dfn}[Influence of a function on the boolean cube]
%Let $f:\{-1,1\}^n \to \R$ be a real function. %The {\em influence of the $i$-th variable} is defined as follows.
%Write $f(x) = x_i f_1(x') + f_2(x')$, where $x'=(x_1,\ldots,x_{i-1},x_{i+1},\ldots,x_n)$. Then
%$$
%\Infi_i(f) = \frac{\E_{x'}[f_1(x')^2]}{\E_x[f(x)^2]},
%$$
%The {\em maximal influence} of $f$ is $\Inf(f) = \max_{i} \Infi_i(f)$.
%\end{dfn}
%A simple calculation shows that $\Infi_i(f)$ can be equivalently
%defined as
%$$
%\Infi_i(f) = \frac{\E_{x'}[\Var_{x_i}[f]]}{\E_x[f(x)^2]}.
%$$
%
%For boolean functions $f:\{-1,1\}^n \to \{-1,1\}$, it is equivalent to
%$$
%\Infi_i(f) = \Pr[f(x) \ne f(x \oplus e_i)]
%$$
%where $e_i$ is the $i$-th unit vector.

\subsection{Analytical tools}
\label{subsection:AnalyticalTools}

The Lipschitz constant of a function bounds the change in the
function value when the inputs are perturbed. It will be
convenient for us to measure distance in the $L_{\infty}$ norm.
Recall that for $z=(z_1,\ldots,z_m) \in \R^m$, its $L_{\infty}$ norm is defined
as the maximal absolute value of its coordinates, i.e.
$$
\|z\|_{\infty} = \max \{|z_i|: i \in [m]\}.
$$

\begin{dfn}[Lipschitz constant]
Let $F:\R^m \to \R$ be a function. The Lipschitz constant of $F$,
denoted by $L(F)$, is defined as
$$
L(F) = \sup_{z',z'' \in \R^m} \frac{|F(z')-F(z'')|}{\|z' -
z''\|_{\infty}}.
$$
The function $F$ is said to be {\em Lipschitz} if $L(F)<\infty$.

Let $C$ be a convex subset of $\R^m$. The Lipschitz constant of
$F$ restricted to $C$, denoted $L_C(F)$, is defined as
$$
L_C(F) = \sup_{z',z'' \in C} \frac{|F(z')-F(z'')|}{\|z' -
z''\|_{\infty}}.
$$.
\end{dfn}

We will use restricted Lipschitz constant only for cubes.

\begin{dfn}
The cubic $\eps$-neighborhood of a point $z \in \R^m$ is defined as
$$
\C(z,\eps) = \{z' \in \R^m: \|z - z'\|_{\infty} \le \eps\}.
$$
For a set $S \subset \R^m$, the cube $\eps$-neighborhood of $S$ is
defined as
$$
\C(S,\eps) = \bigcup_{z \in S} \C(z,\eps).
$$
\end{dfn}
%Ido : I think that the following is not needed, and anyway makes this section too long
% Azuma-Hoeffding bound bounds the tail of linear functions under a uniform assignment of variables.
%\begin{lem}\label{lem:hoeffding}
%Let $g:\{-1,1\}^n \to \R$ be a linear function with $\E[g^2]=1$. Then
%$$
%\Pr_{x \in U}[|g(x)| \ge t] \le 2 e^{-t^2/2}.
%$$
%\end{lem}
\subsection{Tail estimates for polynomials}
\label{subsection:tailEstimates}

In this subsection we prove two results about the concentration of
degree-$d$ multilinear polynomials. The first result gives a tail
estimate on the probability that a degree-$d$ polynomial is very
large, and the second result provides a lower bound on the
probability it is concentrated near a certain value. In both results
we apply techniques based on hyper-contractivity~\cite{MOO05}.

\subsubsection{Tail bounds}

We prove in this subsection a general tail estimate on multilinear
polynomials, which holds both under the uniform distribution over
$\{-1,1\}^{n}$ and under the standard multi-normal distribution.
Namely, we show that for any degree-$d$ multilinear polynomial
$f(x_1,\ldots,x_n)$, the probability that $|f(x)| \ge t$ is
bounded by $\mathrm exp(-t^{2/d})$. We observe that this is tight by considering the
polynomial obtained by multilinearizing $f(x) = (x_1 + \ldots +
x_n)^d$. Our main result follows.

\begin{lem}\label{lem:poly_tail_bound}
Let $f(x_1,\ldots,x_n)$ be a multilinear degree-$d$ polynomial
with $\E[f^2]=1$. Then for every $t \ge 1$,
$$
\Pr_{x \in \B}[|f(x)| \ge t] \le 2^{-\tfrac{d}{4} \cdot t^{2/d}}
$$
and
$$
\Pr_{x \in \N}[|f(x)| \ge t] \le 2^{-\tfrac{d}{4} \cdot t^{2/d}}.
$$
\end{lem}

Let $X$ be a real random variable. Denote
$\|X\|_q=(\E[|X|^q])^{1/q}$. Following the notation
from~\cite{MOO05}, we say that $X$ is $(2,q,\eta)$
hyper-contractive if for every $a \in \R$,
$$
\|a + \eta X\|_q \le \|a + X\|_2.
$$

We use the following two theorems from~\cite{MOO05}.

\begin{lem}[Theorem 3.13 in~\cite{MOO05}]\label{lem:hc_boolean_normal}
If $X$ is uniform on $\{-1,1\}$, or a standard normal random
variable $N(0,1)$, then for every $q>=2$, $X$ is $(2, q, \eta)$
hyper-contractive with $\eta = (q-1)^{-1/2}$.
\end{lem}

\begin{lem}[Proposition 3.12 in~\cite{MOO05}]\label{lem:hc_poly}
Let $X$ be $(2,q,\eta)$ hyper-contractive. Let $f(x_1,\ldots,x_n)$
be a multilinear degree-$d$ polynomial. Let $Q=f(X_1,\ldots,X_n)$
where $X_1,\ldots,X_n$ are i.i.d and distributed according to $X$.
Then
$$
\|Q\|_q \le \eta^{-d} \|Q\|_2
$$
\end{lem}

\begin{proof}[Proof of Lemma~\ref{lem:poly_tail_bound}]
Let $X$ be either a uniform random variable over $\{-1,1\}$ or
standard normal random variable $N(0,1)$. Let
$Q=f(X_1,\ldots,X_n)$ where $X_1,\ldots,X_n$ are i.i.d and
distributed according to $X$. In either case we have $\|Q\|_2 =
\E[f^2]^{1/2}=1$. Fix $q \ge 2$ to be determined later. By
Lemma~\ref{lem:hc_boolean_normal}, $X$ is $(2,q,\eta)$ for
$\eta=(q-1)^{-1/2}$. Thus, by Lemma~\ref{lem:hc_poly} we
have
$$
\E_{x \in X^n}[|f(x)|^q] \le (q-1)^{dq/2}.
$$
Thus by Markov's inequality
$$
\Pr_{x \in X^n}[|f(x)| \ge t^{d/2}] \le
\left(\frac{q-1}{t}\right)^{qd/2}.
$$

Since $t \ge 1$ we can set $q = t/2+1$ and get
$$
\Pr_{x \in X^n}[|f(x)| \ge t^{d/2}] \le 2^{-td/4}.
$$

Hence we conclude
$$
\Pr_{x \in X^n}[|f(x)| \ge t] \le 2^{-\tfrac{d}{4} \cdot t^{2/d} }.
$$
\end{proof}

\subsubsection{Concentration lower bounds}

The main result of this subsection is the following lemma.

\begin{lem}\label{lem:poly_small_inf_anti_concentration}
There exist constants $c_1,c_2>0$ such that the following holds.
Let $f(x_1,\ldots,x_n)$ be a polynomial of degree $d$ such that
$\Var[f]=1$. For $\eps>0$ let $\alpha = (c_1 \cdot
\epsilon/d)^d$ and $\tau = (c_2 \cdot \epsilon/d)^{8d}$. If
$\Inf(f) \le \tau$, then for every $t \in \R$,
$$
\Pr_{x \in \B} [|f(x)-t| \le \alpha] \le \epsilon.
$$
\end{lem}

We use the following two theorems.

\begin{lem}[Theorem 2.1 in~\cite{MOO05}]\label{lem:moo}
Let $f(x_1,\ldots,x_n)$ be a multilinear degree $d$ polynomial,
such that $\Inf(f) \le \tau$. Then for every $t \in \R$
$$
|\Pr_{x \in \B}[f(x) \le t] - \Pr_{x \in N}[f(x) \le t]| \le O(d
\tau^{1/8d}).
$$
\end{lem}

The following is an immediate corollary of Theorem 8 in Carbery
and Wright~\cite{CW01}, which is also stated as Corollary 3.23
in~\cite{MOO05}.
\begin{lem}\label{lem:cw} Let $f(x_1,\ldots,x_n)$ be a multilinear degree $d$
polynomial such that $\Var[f]=1$. Then for every $t \in \R$,
$$
\Pr_{x \in N}[|f(x)-t| \le \alpha] \le O(d \alpha^{1/d}).
$$
\end{lem}

\begin{proof}[Proof of
Lemma~\ref{lem:poly_small_inf_anti_concentration}] Let $f$ be a
degree-$d$ polynomial such that $\Inf(f) \le \tau$. By
Lemma~\ref{lem:moo} we have:
$$
\Pr_{x \in \B}[|f(x)-t| \le \alpha] \le \Pr_{x \in N}[|f(x)-t| \le
\alpha] + O(d \tau^{1/8d}).
$$
By Lemma~\ref{lem:cw} we have
$$
\Pr_{x \in \N}[|f(x)-t| \le \alpha] \le O(d \alpha^{1/d})
$$
Combing the two results we get:
$$
\Pr_{x \in \B}[|f(x)-t| \le \alpha] \le O(d \cdot (\tau^{1/8d} +
\alpha^{1/d})).
$$
Setting $\alpha = (c_1 \cdot \epsilon/d)^d$ and $\tau = (c_2 \cdot
\epsilon / d)^{8d}$ for some absolute constants $c_1,c_2>0$ we get
$$
\Pr_{x \in \B}[|f(x)-t| \le \alpha] \le \epsilon.
$$
\end{proof}

\section{The effect of partial assignments}
\label{sec:structure_results}

We prove in this section that functions with many influential variables cannot
be non-trivially approximated by low-degree PTFs. The proof depends on a new
general structural result for polynomials and polynomial threshold functions.
We show that for every such function there exists a small depth decision tree $D$, such
that $f|_{D}$ has low influence with high probability.

\begin{lem}\label{lem:small_inf_by_decision_tree}
Let $f:\{-1,1\}^n \to \R$ be a degree-$d$ polynomial, and let
$h(x) = \sgn(f(x))$. For every $\epsilon,\delta>0$, there exists a
decision tree $D$ of depth at most $2^{e d/\delta} \cdot
\log(1/\epsilon)$, such that
$$
\Pr_{\ell \in L(D)}[\Inf(f|_{\ell}) > \delta] < \epsilon
$$
and
$$
\Pr_{\ell \in L(D)}[\Inf(h|_{\ell}) > \delta'] < \epsilon
$$
for $\delta' = O(d \cdot \delta^{1/8d})$.
\end{lem}

The proof of Lemma~\ref{lem:small_inf_by_decision_tree} appears in Subsection~\ref{subsec:proof:small_inf_by_decision_tree}.

We apply Lemma~\ref{lem:small_inf_by_decision_tree} in order to prove our main result of this section,
that functions with many influential variables cannot be approximated by low-degree PTFs.
We restate Theorem~\ref{thm:intro:ptf_no_approx} for the convenience of the reader.

\begin{thm}[Theorem~\ref{thm:intro:ptf_no_approx}, restated]\label{thm:noise_sensitive_no_approx}
Let $g:\{-1,1\}^n \to \{-1,1\}$ be a boolean function, such that
$\Infi_i(g) \ge \tau$ for at least $n^{\alpha}$ variables. Then for any degree-$d$
polynomial threshold function $h$ we have
$$
\Pr_{x}[h(x)=g(x)] \le 1 - \frac{\tau}{2} + \eta
$$
where $\eta = O(d / (\alpha \log{n})^{1/8d})$.
\end{thm}

Before proving Theorem~\ref{thm:noise_sensitive_no_approx}, we give a couple of examples for its application.
We show that low-degree PTFs do not admit a non-trivial approximation for the $\MOD_m$ function, or low degree polynomials over $\mathbb F_2$.

\begin{cor}[Corollary~\ref{cor:intro:modm_no_approx}, restated]\label{cor:modm_no_approx}
Let $h:\{-1,1\}^n \to \{-1,1\}$ be a degree-$d$ polynomial
threshold function for $d = O(\log\log{n} / \log\log\log{n})$.
Then
$$
\Pr[h(x) = \MOD_m(x)] \le 1 - \frac{1}{m} + o(1).
$$
%where the $\MOD_M$ function is is defined as
%$$
%\MOD_m(x_1,\ldots,x_n) = \bigg\{
%\begin{array}{cc}
%  1 &  x_1+\ldots+x_n  \equiv 0 \pmod m\\
%  -1 &x_1+\ldots+x_n   \not\equiv 0 \pmod m\\
%\end{array}
%$$
\end{cor}

\begin{proof}
It is straightforward to verify that $\Infi_i(\MOD_m)=\frac{2}{m}$ for all $i \in [n]$, the proof now follows by Theorem~\ref{thm:noise_sensitive_no_approx}.
\end{proof}

\begin{cor}[Corollary~\ref{cor:intro:poly_no_approx}, restated]\label{cor:poly_no_approx}
Let $q:\{-1,1\}^n \to \{-1,1\}$ be a degree-$r$ polynomial over $\F_2$ depending on all variables.
Let $h:\{-1,1\}^n \to \{-1,1\}$ be a degree-$d$ polynomial
threshold function for $d \le O(\log\log n / \log\log\log n)$. Then
$$
\Pr[h(x) = q(x)] \le 1 - 2^{-r} + o(1).
$$
\end{cor}

\begin{proof}
We will prove $\Infi_i(q) \ge 2^{1-r}$ for all $i \in [n]$.
Let $q(x) = (-1)^{q'(x')}$, where $q':\F_2^n \to \F_2$ and $x' \in \F_2^n$ set by $x_i = (-1)^{x'_i}$.
We will in fact show that $\Pr[q'(x') \ne q'(x' \oplus e_i)] \ge 2^{1-r}$.
write $q'(x') = x'_i q'_1(x') + q'_2(x')$. As $q'_1$ is a non-zero polynomial of degree at most $r-1$, we have $\Pr[q'_1(x')=1] \ge 2^{1-r}$.
\end{proof}

We now return to prove Theorem~\ref{thm:noise_sensitive_no_approx}.

\begin{proof}[Proof of Theorem~\ref{thm:noise_sensitive_no_approx}]
Let $g:\{-1,1\}^n \to
\{-1,1\}$ be a boolean function for which $\Infi_{i}(g) \ge \tau$
for at least $n' = n^{\alpha}$ variables. We will provide a lower bound on $q = \Pr[g(x) \ne h(x)]$,

Set $\delta>0$ and $\eps>0$ to be determined later. Set $m=2^{ed /
\delta} \log{1/\eps}$ and $\delta' = O(d \cdot \delta^{1/8d})$.
Using Lemma~\ref{lem:small_inf_by_decision_tree} we get that
there exists a decision tree $D$ of depth at most $m$, such that
$$
\Pr_{\ell \in L(D)}[\Inf(h|_{\ell}) > \delta'] < \eps.
$$

In each path in $D$ there are at most $m$ variables.
Thus, there exists a variable $x_i$ for which $\Infi_i(g) \ge \tau$ which appears in at most $m/n'$
of the paths. Equivalently, a random leaf $\ell \in L(D)$ assigns
a value to $x_i$ with probability at most $m/n'$. We get
\begin{align*}
\Pr[g(x) \ne g(x \oplus e_i)] & \le
\E_{\ell \in L(D)}\left[\Pr_{x}[g|_{\ell}(x) \ne g|_{\ell}(x \oplus e_i)]\right] + m/n' \\
& \le \E_{\ell \in L(D)}[\Pr_{x}[g|_{\ell}(x) \ne
h|_{\ell}(x)]+\Pr_{x}[h|_{\ell}(x) \ne h|_{\ell}(x \oplus e_i)]\\
&+\Pr_{x}[h|_{\ell}(x \oplus e_i) \ne g|_{\ell}(x \oplus e_i)]] +
m/n' \\
& = 2 \Pr[g(x) \ne h(x)] + \E_{\ell \in L(D)}[\Infi_i(h|_{\ell})]
+ m/n' \\
& = 2q + \delta' + \eps + m/n'
\end{align*}

On the other hand, by assumption we have $\Pr[g(x) \ne g(x \oplus
e_i)] \ge \tau$. Combining the two bounds we get that
\begin{align*}
\Pr[g(x) \ne h(x)] & =q \ge \frac{1}{2}(\tau - \eps - \delta' -
m/n') \\
& \ge \frac{\tau}{2} - O(\eps + d \delta^{1/8d} + 2^{ed/\delta}
\log(1/\eps)/n')
\end{align*}
Setting $\delta = O(d/\log{n'})$ and $\eps$ small enough (for
example $\eps=1/n'$) gives
$$
q = \Pr[g(x) \ne h(x)] \ge \frac{\tau}{2} - \eta
$$
for $\eta = O(\frac{d}{(\alpha \log{n})^{1/8d}})$.
\end{proof}

\subsection{Proof of Lemma~\ref{lem:small_inf_by_decision_tree}}
\label{subsec:proof:small_inf_by_decision_tree}

The proof of Lemma~\ref{lem:small_inf_by_decision_tree} will be
conducted in three steps. First we show that for every low-degree
polynomial there exists a partial assignment of a small set of
variables under which we get a polynomial with low influences. We
then argue that if a polynomial has low influences, then so does
its threshold. We then conclude by showing that if there is a single good assignment, then by taking larger set of variables we get that most of the assignments are good. The first step is accomplished by the
following lemma.

\begin{lem}\label{lem:small_inf_by_assignment}
Let $f:\{-1,1\}^n \to \R$ be a degree-$d$ polynomial. For every
$\delta>0$ there exist a set of variables $x_{i_1},\ldots,x_{i_k}$
and assignments for these variables $b_{i_1},\ldots,b_{i_k} \in
\{-1,1\}$, such that
$$
\Inf(f|_{x_{i_1}=b_{i_1},\ldots,x_{i_k}=b_{i_k}}) \le \delta
$$
and $k \le ed/\delta$.
\end{lem}

\begin{proof}
We construct a sequence of assignments for the variables of $f$, assigning a value to
a single variable at each step, that will lead eventually to a
polynomial $f|_{x_{i_1}=b_{i_1},\ldots,x_{i_k}=b_{i_k}}$ whose
influence is bounded by $\delta$.

Every degree-$d$ polynomial $f$ can be uniquely represented as
$$
f(x) = \sum_{I \subset [n], |I| \le d} f_I \prod_{i \in I} x_i.
$$

For $\alpha \ge 0$ define operator $V_{\alpha}(f)$ to be$$
V_{\alpha}(f) = \sum_{I \subset [n], |I| \le d} |f_I|^2
(1+\alpha)^{|I|}.
$$
Note that $V_0(f) = \E[f^2]$.

Fix a variable $x_i$, and let $f(x) = x_i f_1(x') + f_2(x')$
where $x'=(x_1,\ldots,x_{i-1},x_{i+1},\ldots,x_n)$. We have
$f|_{x_i=1} = f_1+f_2$ and $f|_{x_i=-1} = -f_1+f_2$. Notice that
$V_0(f_1) = Inf_i(f) \cdot V_0(f)$.

We first claim that
\begin{align}\label{eq:valpha_avg}
\tfrac{1}{2}(V_{\alpha}(f|_{x_i=1})+V_{\alpha}(f|_{x_i=-1})) =
V_{\alpha}(f) - \alpha V_{\alpha}(f_1)
\end{align}

To prove it, write $f_1(x') = \sum f_{1,I} \prod_{i \in I} x'_i$ and $f_2(x') =
\sum f_{2,I} \prod_{i \in I} x'_i.$ We have

\begin{align*}
& V_{\alpha}(f|_{x_i=1})+V_{\alpha}(f|_{x_i=-1}) = V_{\alpha}(f_1 + f_2)+V_{\alpha}(-f_1+f_2) = \\
& \sum_I (f_{1,I} + f_{2,I})^2 (1+\alpha)^{|I|}+ \sum_I (-f_{1,I} + f_{2,I})^2 (1+\alpha)^{|I|}= \\
& 2 \cdot \sum_I(f_{1,I}^2 + f_{2,I}^2) (1+\alpha)^{|I|} = \\
& 2 \cdot \sum_I(f_{1,I}^2 (1+\alpha)^{|I|+1} + f_{2,I}^2 (1+\alpha)^{|I|}) -  2\alpha \cdot \sum_I f_{1,I}^2 (1+\alpha)^{|I|}= \\
& 2 \cdot (V_{\alpha}(f)-\alpha V_{\alpha}(f_1))
\end{align*}

This proves~\eqref{eq:valpha_avg}. In particular for $\alpha = 0$
we get
\begin{align}\label{eq:v0_avg}
\tfrac{1}{2}(V_0(f|_{x_i=1})+V_0(f|_{x_i=-1})) = V_0(f).
\end{align}

and for $\alpha>0$ we have
\begin{align}\label{eq:valpha_inf}
\tfrac{1}{2}(V_{\alpha}(f|_{x_i=1})+V_{\alpha}(f|_{x_i=-1})) \le
V_{\alpha}(f) - \alpha \cdot Inf_i(f) \cdot V_0(f),
\end{align}
since $V_{\alpha}(f_1) \ge V_0(f_1) = Inf_i(f) \cdot V_0(f)$.

%Since $V_{\alpha}(f_1) \ge V_0(f_1) = Inf_i(f)$, we also get
%\begin{align}\label{eq:v0_avg}
%\tfrac{1}{2}(V_{\alpha}(f|_{x_i=1})+V_{\alpha}(f|_{x_i=-1})) \le V_{\alpha}(f) - Inf_i(f)
%\end{align}

Define $S_{\alpha}(f) = \frac{V_{\alpha}(f)}{V_{0}(f)}$. We next
prove that
\begin{align}\label{eq:s_avg}
\min\sr{S_{\alpha}(f|_{x_i=1}), S_{\alpha}(f|_{x_i=-1})} \le
S_{\alpha}(f) - \alpha \cdot Inf_i(f)
\end{align}

By combining ~\eqref{eq:valpha_avg} and~\eqref{eq:v0_avg} we
get
\begin{align}\label{eq:v_avg} S_{\alpha}(f) =
&\frac{V_{\alpha}(f)}{V_0(f)} = \frac{ V_{\alpha}(f|_{x_i=1}) +
V_{\alpha}(f|_{x_i=-1})}{ V_{0}(f|_{x_i=1}) + V_{0}(f|_{x_i=-1})}+
\frac{\alpha V_{\alpha}(f_1)}{V_0(f)} \ge \\
&\min\sr{\frac{ V_{\alpha}(f|_{x_i=1})}{V_{0}(f|_{x_i=1})},\frac{
V_{\alpha}(f|_{x_i=-1})}{V_{0}(f|_{x_i=-1})}} +
\frac{\alpha V_{0}(f_1)}{V_0(f)} = \\
&\min\sr{S_{\alpha}(f|_{x_i=1}), S_{\alpha}(f|_{x_i=-1})} + \alpha
\cdot Inf_i(f)
\end{align}

Consider the polynomial $f$. We first bound $S_{\alpha}(f)$,
$$
S_{\alpha}(f)=\frac{V_{\alpha}(f)}{V_0(f)} = \frac{\sum_I |f_I|^2
(1+\alpha)^{|I|}}{\sum_I |f_I|^2}\le (1+\alpha)^d.
$$

Note that either $\Inf(f)\le \delta$, or there exists a variable $x_{i_1}$,
such that
$$
\min\sr{S_{\alpha}(f|_{x_{i_1}=1}),S_{\alpha}(f|_{x_{i_1}=-1})}\le
S_{\alpha}(f)-\alpha \cdot \delta
$$
Consider the restriction $f_{x_{i_1}=b_{i_1}}$ for $b_{i_1} \in
\{-1,1\}$ minimizing $S_{\alpha}(f_{x_{i_1}=b_{i_1}})$. Either
$\Inf(f_{x_{i_1}=b_{i_1}}) \le \delta$, or otherwise we could find
another variable $x_{i_2}$ such that
$$
\min\sr{S_{\alpha}(f|_{x_{i_1}=b_{i_1},
x_{i_2}=1}),S_{\alpha}(f|_{x_{i_1}=b_{i_1},x_{i_2}=-1})}\le
S_{\alpha}(f|_{x_{i_1}=b_{i_1}})-\alpha \cdot \delta
$$
Continuing in this fashion, since $S_{\alpha} \ge 0$, we must
reach after at most $k \le \frac{(1+\alpha)^d}{\alpha \delta}$
steps a polynomial $f|_{x_{i_1}=b_{i_1},\ldots,x_{i_k}=b_{i_k}}$
such that $\Inf(f|_{x_{i_1}=b_{i_1},\ldots,x_{i_k}=b_{i_k}}) \le
\delta$. Choosing optimally $\alpha=\frac{1}{d-1}$ we get $k \le e
\cdot d/\delta$.
\end{proof}

We now show that if a polynomial has low influences, then so does
its threshold.

\begin{lem}\label{lem:poly_low_inf_PTF_low_inf}
Let $f:\{-1,1\}^n \to \R$ be a degree-$d$ polynomial such that
$\Inf(f)=\delta$. Let $h(x) = \sgn(f(x))$. Then
$$
\Inf(h) \le O(d \cdot \delta^{1/8d}).
$$
\end{lem}

\begin{proof}
Assume w.l.o.g $\Var[f]=1$, and we will bound $\Infi_i(h)$ for all
$i=1,\ldots,n$.

We first argue that if $\E[f^2]$ is large, then $h$ has low
influences. Let $f(x) = c + f_0(x)$, where $c$ is the free
coefficient of $f$. We have $\Var[f] = \E[f_0^2]=1$ and $\E[f^2] =
1+c^2$. The probability that $h(x)=h(0)$ is bounded by
$$
\Pr[h(x)=h(0)] \le \Pr[|f_0(x)| \ge c] \le
\frac{\E[f_0^2]}{c^2}=\frac{1}{c^2}.
$$
Thus for large $c$ we get a bound on the influence of $h$, since
$$
\Infi_i(h) = \Pr[h(x) \ne h(x \oplus e_i)] \le \Pr[h(x) \ne h(0)]
+ \Pr[h(x \oplus e_i) \ne h(0)] \le 2/c^2.
$$

In particular if $c > \delta^{-1/4}$ we get that $\Infi_i(h) \le
O(\delta^{1/2})$ and we are done. We thus assume from now on that $c
\le \delta^{-1/4}$.

Let $f(x) = x_i f_1(x) + f_2(x)$, where $f_1,f_2$ do not depend on
$x_i$. By our assumption on the influences,
$$
\E_x[f_1^2] = \Infi_i(f) \cdot \E[f^2] \le \delta(1+c^2) \le 2
\delta^{1/2}.
$$

Set $a = \delta^{1/8}$ and consider the following two cases.
\begin{enumerate}
    \item $|f(x)| \le a$
    \item $|f_1(x)| \ge a$
\end{enumerate}

If neither of these cases occur, then flipping $x_i$ does not
change the sign of $f$. Thus we can bound
$$
\Infi_i(h) \le \Pr[|f(x)| \le a] + \Pr[|f_1(x)| \ge a].
$$

We first estimate the first summand. By Lemma~\ref{lem:poly_small_inf_anti_concentration}. Set
$\tilde{\delta} \ge \max(\tfrac{d}{c_1} a^{1/d}, \tfrac{d}{c_2}
\delta^{1/8d})$ where $c_1,c_2$ are the constants in
Lemma~\ref{lem:poly_small_inf_anti_concentration}. We get
$$
\Pr[|f(x)| \le a] \le \tilde{\delta} = O(d \cdot \delta^{1/8d}).
$$

We proceed by estimating the second summand. By Markov inequality and get
$$
\Pr[|f_1(x)| \ge a] \le \frac{E[f_1^2]}{a^2} \le 2 \delta^{1/4}.
$$

Combining the two estimations we get that
$$
\Infi_i(h) \le O(d \cdot \delta^{1/8d}),
$$
as desired.

\end{proof}

We next prove Lemma~\ref{lem:small_inf_by_decision_tree}. Using
Lemma~\ref{lem:small_inf_by_assignment} we prove the existence of
a small depth decision tree, such that for most of its leaves, the
polynomial restricted to the leaf has low influences. We use
Lemma~\ref{lem:poly_low_inf_PTF_low_inf} to argue that when this
happens also the threshold function has low influences.

\begin{proof}[Proof of Lemma~\ref{lem:small_inf_by_decision_tree}]
We first prove the theorem for a polynomial $f$, and then for a PTF h. We build a
decision tree $D$ in steps. At every step, some of the leaves of
$D$ will be {\em open}, and some will be {\em closed}. If a leaf
${\ell}$ is closed then $\Inf(f|_{\ell}) \le \delta$. A leaf is
open if it is not closed. Initially, our tree consists a single vertex, the root, which is open.

Let $\ell$ be an open leaf, and consider the polynomial
$f|_{\ell}$. By Lemma~\ref{lem:small_inf_by_assignment}, there
exist a set of variables $x_{i_1},\ldots,x_{i_k}$, $k \le \frac{ed}{\delta}$ and an
assignment to these variables $b_{i_1},\ldots,b_{i_k} \in
\{-1,1\}$, such that
$$
\Inf(f|_{\ell,x_{i_1}=b_{i_1},\ldots,x_{i_k}=b_{i_k}}) \le \delta.
$$

We add under a $\ell$ a subtree whose leaves correspond to all the $2^{k}$ possible assignments of $x_{i_1},\ldots,x_{i_k}$.
Note that at least one of the leaves in the new tree is closed, and the other leaves may be either closed or open. Therefore, a random walk of length $k$ that starts at $\ell$ will end at a closed leaf with probability at least $2^{-k}$.

This process defines a tree $D'$ of depth at most $n$, as every variable appears in every path at most once. Let $D(t)$ be the tree obtained by truncating $D'$ at depth $t \cdot 2^k$. Namely, the depth of $D(t)$ is $t \cdot 2^k$. The probability that a random walk that start from the root will end at open leaf is at most $(1-2^{-k})^t \le e^{-2^{-k} t}$. Thus, setting, $t = \log(1/\epsilon) \cdot 2^{e d/\delta}$ will
guarantee that a random leaf in $D$ is closed with probability at
least $1-\epsilon$, as required.

We proceed by proving the second item. Let $h$ be a PTF as stated, and observe that
by Lemma~\ref{lem:poly_low_inf_PTF_low_inf}, for any leaf $\ell$
for which $\Inf(f|_{\ell}) \le \delta$ we have that
$\Inf(\sgn(f|_{\ell})) \le O(d \delta^{1/8d}) = \delta'$. Since
$\sgn(f|_{\ell}) = \sgn(f)|_{\ell} = h|_{\ell}$, we get
$$
\Pr_{\ell \in L(D)}[\Inf(h|_{\ell}) > \delta'] < \eps.
$$
\end{proof}

\section{Fooling threshold of polynomials depending on a few linear functions}
\label{section-approx-and-fool}

Recall that the weight of a polynomial $G:\R^m \to \R$ is the sum of the absolute
values of the coefficients of its monomials, excluding the free coefficient. Our main result in this section is Theorem~\ref{thm:kwise_fool_poly_small_coefs}, which is stated below.

\begin{thm}[Theorem~\ref{thm:intro:prg_poly_few_linear}, restated]\label{thm:kwise_fool_poly_small_coefs}
Fix $\eps>0$. Let $f:\{-1,1\}^n \to \R$ be a degree-$d$ polynomial,
which can be decomposed as $f(x) = G(g_1(x),\ldots,g_m(x))$ where
\begin{enumerate}
    \item The functions $g_1,\ldots,g_m$ are linear with $\E[g_1^2]=\ldots=\E[g_m^2]=1$.
    \item $G$ is a degree-$d$ polynomial.
\end{enumerate}
Then $k$-wise distributions $\eps$-fool $\sgn(f)$ for $k = \mathrm exp(O(d/\eps)^d) + poly((\log{m} \cdot d/\eps)^d, m, wt(G))$.
\end{thm}

The main lemma shows that any multivariate Lipschitz function $F$ admits a polynomial $p$ with the following two properties. The polynomial $p$ bounds $F$ from above everywhere, and $p$ approximates $F$ in a cube around the origin.
\begin{lem}\label{lem:approx_and_bound}
Let $F:\R^m \to [-1,1]$ be a Lipschitz function. Let $A>0$ and
$0<\eps<1$ be arbitrary. There exists a degree-$k$ polynomial
$p(z_1,\ldots,z_m)$ such that
\begin{enumerate}
  \item For every $z \in \R^m$, $p(z) \ge F(z)$.
  \item For every $z \in [-A,A]^m$, $p(z) \le F(z) + \eps$.
\end{enumerate}
and $k \le O\left(\frac{A \cdot m^{3/2} \cdot L(F)}{\epsilon}\right)$.
\end{lem}

The proof of Lemma~\ref{lem:approx_and_bound} appears in Subsection~\ref{}

We next apply Lemma~\ref{lem:approx_and_bound} to show that $k$-wise distributions fool any boolean function $f:\{-1,1\}^n \to \R$ with the following properties. The function $f$ be decomposed as $f(x) = G(g_1(x),\ldots,g_m(x))$, where $g_1,\ldots,g_m$ are linear functions, the polynomial $G$ is Lipschitz, and the distribution of $f$ is not too concentrated around any specific value.

\begin{lem}\label{lem:approx_thresh}
Let $f:\{-1,1\}^n \to \R$ be a function which can be decomposed as
$f(x) = G(g_1(x),\ldots,g_m(x))$ where
\begin{enumerate}
    \item The functions $g_1,\ldots,g_m:\{-1,1\}^n \to \R$ are linear with $\E[g_1^2],\ldots,\E[g_m^2] \le 1$.
    \item The function $G:\R^m \to \R$ is continuous and Lipschitz on the cube $[-C,C]^{m}$, for $C = 100
\sqrt{\log(m/\eps)}$.
    \item The function $f$ is anti-concentrated, $\Pr_{x}[|f(x)| <\alpha] < \eps/100$ for some $\alpha$ depending on $\eps$.
\end{enumerate}
Then there exists a degree-$k$ polynomial $p:\{-1,1\}^n \to \R$ such that
\begin{itemize}
  \item $p(x) \ge \sgn(f(x))$ for all $x \in \{-1,1\}^n$.
  \item $\E_{x \in U}[p(x) - \sgn(f(x))] \le \eps$.
\end{itemize}
where $k=O(\frac{d m^5 L^2}{\alpha^2 \eps^2} \ln(mL/\alpha
\eps))$ and $L = \max(L_{[-C,C]^m}(g),1)$.
\end{lem}

The following claim bounds the Lipschitz constant of degree-$d$ polynomials.

\begin{clm}\label{clm:lip_of_poly_bounded_wt}
Let $G:\R^m \to \R$ be a degree-$d$ polynomial. The Lipschitz constant of $G$
on $[-C,C]^m$ is bounded by $d C^{d-1} \cdot wt(G)$.
\end{clm}
\begin{proof}
We start by bounding the Lipschitz constant of monomials on $[-C,C]^m$. We then
will get the result for $G$ by the additivity of the Lipschitz constant.

Let $M$ be a monomial $M(z_1,\ldots,z_d) = \prod z_i$. Let $z,z' \in [-C,C]^m$ such that $\|z-z'\|_{\infty} \le \eps$.
Let $z'_i = z_i + e_i$ where $|e_i| \le \eps$. We have
\begin{align*}
|M(z') - M(z)| & = |\sum_{k=1}^d \left( \prod_{i=1}^k (z_i + e_i) \prod_{i=k+1}^d z_i -
\prod_{i=1}^{k-1} (z_i + e_i) \prod_{i=k}^d z_i\right)| \le \\
& \sum_{k=1}^d \prod_{i=1}^{k-1} |z_i| \prod_{i=k+1}^d |z_i+e_i| e_i \le \\
& d C^{d-1} \eps.
\end{align*}
Hence $L_{[-C,C]^m}(M) \le d C^{d-1}$.

Write $G(z) = \sum_{I \subset [m], |I| \le d} \alpha_I M_I(z)$ where $M_I$ are monomials. The Lipschitz constant
of $G$ on $[-C,C]^m$ is thus bounded by $\sum |\alpha_I| L_{[-C,C]^m}(M_I) \le d C^{d-1} \cdot wt(G)$.
\end{proof}

We proceed to the proof of Theorem~\ref{thm:kwise_fool_poly_small_coefs}.

\begin{proof}[Proof of Theorem~\ref{thm:kwise_fool_poly_small_coefs} ]
Let $f(x)$ be a degree-$d$ polynomial, which can be decomposed as $f(x)=G(g_1(x),\ldots,g_m(x))$ where
$g_1,\ldots,g_m$ are linear and $\E[g_1^2]=\ldots=\E[g_m^2]=1$. Set $\delta = O(\eps/d)^{8d}$. By Lemma~\ref{lem:small_inf_by_decision_tree} there exists a decision tree $D$ of depth at most
$\mathrm exp(O(d^{8d+1}/\eps^{8d})$ such that
$$
\Pr_{\ell \in L(D)}[\Inf(f|_{\ell}) > \delta] < \eps/100.
$$

By Lemma~\ref{lem:L2_linear_decision_tree} we have for each linear function $g_i$
$$
\Pr_{\ell \in L(D)}[\E[((g_i)|_{\ell})^2] \ge t] \le \eps/100m
$$
for $t = O(\log{\eps/m})$. Thus with probability $1 - \eps/100$, we have both that
$\Inf(f|_{\ell}) \le \delta$ and $\E[((g_i)|_{\ell})^2] \le t$ for all $i \in [m]$.
Fix such $\ell$. Since $f|_{\ell}$ has low influences, Lemma~\ref{lem:poly_small_inf_anti_concentration}
gives
$$
\Pr_{x \in U}[|f|_{\ell}(x)| \le \alpha] < \eps/1000.
$$
for $\alpha = O(\eps/d)^d$.

Let $g_i'$ be a normalization of $(g_i)_{\ell}$ such that $\E[(g_i')^2]=1$. We can write
$f_{\ell}(x) = G'(g_1'(x),\ldots,g_m'(x))$ where $wt(G') \le wt(G) \cdot t$. By Claim~\ref{clm:lip_of_poly_bounded_wt}
we have $L_{[-C,C]^m}(G) \le d C^{d-1} \cdot wt(G)$ for $C=100 \sqrt{\log(m/100\eps)}$. Applying Lemma~\ref{lem:approx_thresh} we get
there exists a degree-$k$ polynomial $p_u(x)$ such that both $p_u(x) \ge \sgn(f|_{\ell}(x))$ for all $x \in \{-1,1\}^n$, and $\E_{x \in U}[p_u(x) - \sgn(f|_{\ell}(x))] \le \eps/10$. Applying the same reasoning on the polynomial $-f(x)$
we get there exists a degree-$k$ polynomial $p_l(x)$ such that both $p_l(x) \le \sgn(f|_{\ell}(x))$ for all $x \in \{-1,1\}^n$
and $\E_{x \in U}[\sgn(f|_{\ell}(x)) - p_l(x)] \le \eps/10$. Combining the two bounds we conclude that $k$-wise
distributions $\eps/10$-fool $f|_{\ell}$. Since this holds for $1 - \eps/100$ fraction of the leaves $\ell$,
we get by Claim~\ref{clm:enough_fool_leaves} that $k'=k+\depth(D)$ independence $\eps$-fool $f$.

We conclude by bounding $k$ and $k'$. We have $k = O(\frac{d m^5 L^2}{\alpha^2 \eps^2} \log(mL/\alpha \eps)) =
O(d/\eps)^{d} \cdot m^5 wt(G)^2 \log(m/\eps)^d \cdot O(\log(d \cdot m \cdot wt(G) / \eps))$, and $\depth(D) = \mathrm exp((d/\eps)^{O(d)}$,
hence we have $k' = \mathrm exp((d/\eps)^{O(d)} + \mathrm poly(O(\log{m} \cdot d/\eps)^{d}, m, wt(G))$, as claimed.
\end{proof}

\subsection{Proof of Lemma~\ref{lem:approx_and_bound}}
\label{subsec:proof:approx_and_bound}

Our starting point is a fundamental result in
the theory of approximation theory. Roughly speaking, it says that
any Lipschitz function can be well approximated by a low-degree
polynomial on a bounded region. Explicitly we use the following result of Ganzburg~\cite{Gan79}.

\begin{lem}[Multidimensional Jackson-type theorem, Theorem~1 in~\cite{Gan79}]\label{lem:md_jackson}
Let $F:\R^m \to \R$ be a Lipschitz function. For every $k$ there is
a degree-$k$ polynomial $p_k(z_1,\ldots,z_m)$, such that
$$
\sup_{z \in [-1,1]^m} |F(z) - p_k(z)| \le C \cdot \frac{m^{3/2} L(F)}{k}
$$
where $C$ is an absolute constant.
\end{lem}

We get the following corollary.

\begin{cor}\label{cor:jackson_bounded_cube}
Let $F:\R^m \to \R$ be a Lipschitz function. For every $\eps>0$
there exists $k = O(m^{3/2} L(F) / \eps)$ and a degree $k$ polynomial $p_k$ such that
\begin{itemize}
    \item $p_k(z) \ge F(z)$ for all $z \in [-1,1]^m$
    \item $p_k(z) - F(z) \le \eps$ for all $z \in [-1,1]^m$.
\end{itemize}
\end{cor}

\begin{proof}
Let $p_k$ be the polynomial obtained by Lemma~\ref{lem:md_jackson} such that
$\sup_{z \in [-1,1]^m} |F(z) - p_k(z)| < \eps/2$, and take $p'_k(z) = p_k(z)+\eps/2$.
\end{proof}

We also need the following bound on the growth of real polynomials.

\begin{lem}\label{lem:dubiner}
let $g(w)$ be a univariate degree-$k$ polynomial. Then for every
$w \in \R$,
$$
|g(w)| \le (\max_{w \in [-1,1]} |g(w)|) \cdot |w +
\sqrt{w^2-1}|^k.
$$
\end{lem}

We will need the following corollary of Lemma~\ref{lem:dubiner}.

\begin{lem}\label{lem:dubiner_md}
Let $p(z_1,\ldots,z_m)$ be a degree-$k$ polynomial, such that
$p(z) \le c$ for all $z \in [-1,1]^m$. If $|z_i| \ge |z_j|$ for every $1 \le i,j \le m$, then
$$
|p(z)| \le c \cdot \max(|2 z_i|^k, 1).
$$
\end{lem}

\begin{proof}
Assume w.l.o.g that $|z_1| \ge |z_i|$ for every $i \in \{1,\ldots,m\}$. If
$|z_1| \le 1$ that $(z_1,\ldots,z_m) \in [-1,1]^m$ and by
assumption $p(z) \le c$. Otherwise consider the following univariate polynomial
$g(w)$ that is obtained by restricting $p$ to the line passing through
zero and $z$, defined as
$$
g(w) = p(w, w z_2/z_1, \ldots, w z_m/z_1).
$$

When $w \in [-1,1]$, we have $(w, w z_2/z_1, \ldots, w z_m/z_1)
\in [-1,1]^m$. Hence $\max_{w \in [-1,1]}g(w) \le c$. Applying
Lemma~\ref{lem:dubiner} we get that
$$
|p(z)| = |g(z_1)| \le c \cdot |z_1 + \sqrt{z_1^2-1}|^k \le c \cdot
|2 z_1|^k.
$$
\end{proof}

We are now ready to state and prove the main lemma that will be used
to prove Lemma~\ref{lem:approx_and_bound}.

\begin{lem}\label{lem:jackson_global_bound}
Let $F:\R^m \to [-1,1]$ be a Lipschitz function.
For every $0<\eps<1$ there exists a degree-$k$ polynomial $p'_k$ such that
\begin{itemize}
    \item $p'_k(z) \ge F(z)$ for all $z \in \R^m$.
    \item $p'_k(z) - F(z) \le \eps$ for all $z \in [-1/4,1/4]^m$.
\end{itemize}
where $k = O(m^{3/2} L(F) /\eps)$.
\end{lem}

\begin{proof}
Let $p_k$ be the polynomial guaranteed by Corollary~\ref{cor:jackson_bounded_cube} for error
$\eps/2$. Set $k' \ge \max(k, 4m/\eps)$ be an even integer, and define
$$
p'_k(z_1,\ldots,z_m) = p_k(z_1,\ldots,z_m) +
4\left((2x_1)^{k'} + \ldots + (2 x_m)^{k'}\right).
$$
We will prove that $p'_k(z) \ge F(z)$ for all $z \in \R^m$, and
$p'_k(z) \le F(z)+\eps$ for $z \in [-1/4,1/4]^m$.

Let $z \in \R^m$ be arbitrary. If $z \in [-1,1]^m$ we already have
that $p'_k(z) \ge p_k(z) \ge F(z)$. Otherwise, assume w.l.o.g that
$|z_1| \ge \max(|z_2|,\ldots,|z_m|)$, and hence $|z_1|>1$.

Since $p_k$ approximates $F$ with error $\eps<1$ on $[-1,1]^m$, we
have that $|p_k(z)| \le 2$ for all $z \in [-1,1]^m$. Applying
Lemma~\ref{lem:dubiner_md} we get that
$$
p_k(z) \le 2 |2 z_1|^{k}.
$$

Thus in particular, $p_k(z) \ge -2 |2 z_1|^k$. By our definition
of $p'_k(z)$ we get that
\begin{align*}
p'_k(z) = & p_k(z) + 4\left((2x_1)^{k'} + \ldots + (2 x_m)^{k'}\right)\\
& \ge -2 |2 z_1|^k + 4 ((2z_1)^{k'} + \ldots + (2 z_m)^{k'}) \\
& \ge -2 |2 z_1|^k + 4 (2z_1)^{k'} \\
& = -2 |2 z_1|^k + 4 |2 z_1|^{k'} \\
& \ge -2 |2 z_1|^k + 4 |2 z_1|^k \\
& = 2 |2 z_1|^k \ge 1.
\end{align*}
and in particular we get that
$$
p'_k(z) \ge F(z).
$$

We next estimate the obtained approximation of $p'_k$ in $[-1/4,1/4]^m$.
Observe that for $z \in [-1/4,1/4]^m$,
$$
|p'_k(z) - p_k(z)| \le 4m 2^{-k'}
$$
and by our choice of $k'$, we have that $|p'_k(z) - P_k(z)| \le
\eps/2$. Since $p_k$ approximates $F$ on $[-1,1]^m$ with error $\eps/2$,
it does so in particular in $[-1/4,1/4]^m$. Hence we get
$$
\max_{z \in [-1/4,1/4]^m} p'_k(z) - F(z) \le \eps.
$$
\end{proof}

The proof of Lemma~\ref{lem:approx_and_bound} now follows as an immediate corollary of Lemma~\ref{lem:jackson_global_bound}.

\begin{proof}[Proof of Lemma~\ref{lem:approx_and_bound}]
Let $F:\R^m \to [-1,1]$ be a Lipschitz function. Define $F'(z) = F(z / 4A)$, and apply Lemma~\ref{lem:jackson_global_bound} on $F'$ to obtain a polynomial $p'_k$ such that
$p'_k(z) \ge F'(z)$ for all $z \in \R^m$ and $p'_k(z) \le F'(z) + \eps$ for $z \in [-1/4,1/4]^m$.
The polynomial $p(z) = p'_k(4A \cdot z)$ is the desired approximation polynomial for $F$.
The bound on the degree follows from Lemma~\ref{lem:jackson_global_bound} since
$L(F') = 4A \cdot L(F)$.
\end{proof}

\subsection{Proof of Lemma~\ref{lem:approx_thresh}}
\label{subsec:proof:approx_thresh}

We start with the following definition.

\begin{dfn}[zero-set]
For $G:\R^m \to \R$ we define its zero-set, denoted
$\mathcal{Z}(G)$ to be
$$
\mathcal{Z}(G) = \{z \in \R^m: G(z)=0\}.
$$
\end{dfn}

\begin{lem}\label{lem:lip_approx}
Let $G:\R^m \to \R$ be a continuous real function. For every
$\tau>0$ there exists a function $\tilde{G}:\R^m \to [-1,1]$ such
that
\begin{itemize}
  \item $\tilde{G}(z) \ge \sgn(G(z))$ for all $z \in \R^m$.
  \item For every $z \notin \C(\mathcal{Z}(G),\tau)$, $\tilde{G}(z) = \sgn(G(z))$.
  \item $L(\tilde{G}) \le O(m / \tau)$.
\end{itemize}
\end{lem}

\begin{proof}
Set $\tau' = \tau/2$ and define
$$
G'(z) = \max_{z' \in \C(z,\tau')} \sgn(G(z))
$$
and
$$
\tilde{G}(z) = \frac{1}{|\C(z,\tau')|} \int_{z' \in \C(z,\tau')}
G'(z') dz'.
$$

First we argue that $\tilde{G}(z) \ge \sgn(G(z))$ for all $z \in
\R^m$. Since for every $z' \in \C(z,\tau')$, $G'(z') \ge
\sgn(G(z))$. By definition, $\tilde{G}(z)$ is defined as the average
of $G'(z')$ over $z' \in \C(z,\tau')$, we get that $\tilde{G}(z)
\ge \sgn(G(z))$.

We continue by showing that $\tilde{G}(z) = \sgn(G(z))$ for $z \notin
\C(\mathcal{Z}(G), \tau)$. For $z' \in \C(z,\tau)$ we have
$\sgn(G(z'))=\sgn(G(z))$, since $G$ is continuous and has no zeros
in $\C(z,\tau)$. As $\C(z',\tau') \subset \C(z, \tau)$, we have
$G'(z')=\sgn(G(z))$, and hence we conclude that
$\tilde{G}(z)=\sgn(G(z))$.

We next bound $L(G)$. Let $z',z'' \in \R^m$. We consider the
following two cases. If $\|z' - z''\|_{\infty} \ge \tau'$ then
since $\tilde{G}$ is bounded, i.e. $|\tilde{G}|_{\infty} \le 1$,
we have
$$
\frac{|\tilde{G}(z')-\tilde{G}(z'')|}{\|z'-z''\|_{\infty}} \le
2/\tau'.
$$

Otherwise, if $\|z' - z''\|_{\infty} < \tau'$, we have
\begin{align*}
|\tilde{G}(z')-\tilde{G}(z'')| = &
 |\frac{1}{(2 \tau')^m} \left( \int_{t \in \C(z',\tau')} G'(t) dt - \int_{t \in \C(z'',\tau')} G'(t) dt \right)| \le \\
& \frac{1}{\tau^m} \int_{t \in \C(z',\tau') \triangle \C(z'',\tau')} |G'(t)| dt \le \\
& \frac{|\C(z',\tau') \triangle \C(z'',\tau')|}{\tau^m}
\end{align*}
where $\triangle$ denotes the symmetric difference between two
sets.

A straight forward calculation shows that
$$
|\C(z',\tau') \triangle \C(z'',\tau')| \le O(m (2\tau')^{m-1}
\|z'-z''\|_{\infty}).
$$

Hence we get
$$
\frac{|\tilde{G}(z')-\tilde{G}(z'')|}{\|z'-z''\|_{\infty}} \le O(m
/\tau').
$$
\end{proof}

\begin{lem}\label{lem:zeros_nbd_small_prob}
Let $f(x)=G(g_1(x),\ldots,g_m(x))$ as in the definition of
Lemma~\ref{lem:approx_thresh} and assume that the assumptions of Lemma~\ref{lem:approx_thresh} hold. Then
$$
\Pr_{x \in \{-1,1\}^n}[(g_1(x),\ldots,g_m(x)) \in
\C(\mathcal{Z}(G),\tau)] \le \eps/10
$$
for $\tau = \alpha/L$.
\end{lem}

\begin{proof}
We consider two cases, the first when $(g_1(x),\ldots,g_m(x)) \in
[-(C-\tau),C-\tau]^m$, and the second when $(g_1(x),\ldots,g_m(x))
\notin [-(C-\tau),C-\tau]^m$.

In the first case, let $x \in \{-1,1\}^n$ be such that
$(g_1(x),\ldots,g_m(x)) \in [-(C-\tau),C-\tau]^m \bigcap
\C(\mathcal{Z}(G),\tau)$. We will prove that $|f(x)|<\alpha$, and
by our assumption the probability over all $\{-1,1\}^n$ that
$|f(x)|<\alpha$ is bounded by $\eps/10$. To show that
$|f(x)|<\alpha$, let $z=(g_1(x),\ldots,g_m(x)) \in \R^m$. $z$ is
in $L_{\infty}$ distance of at most $\tau$ from a zero $z_0$ of
$G$, and since $z \in [-(C-\tau),C-\tau]^m$, we get that $z_0 \in
[-C,C]^m$. Since $G$ is Lipschitz on $[-C,C]^m$, we conclude that
$$
G(z) \le G(z_0) + L_{[-C,C]^m} \cdot \|z - z_0\|_{\infty} \le
L_{[-C,C]^m} \cdot \tau \le \alpha.
$$

We now consider the second case, that $(g_1(x),\ldots,g_m(x))
\notin [-(C-\tau),C-\tau]^m$. We will bound the probability that
this event occurs. By our construction $\tau \le 1$, hence it is
enough to bound the probability that $(g_1(x),\ldots,g_m(x))
\notin [-(C-1),C-1]^m$, i.e. $|g_i(x)| \ge C-1$ for some $i \in
[m]$. Since we assumed each $g_i$ is $\delta$-normal, we get that
$$
\Pr[|g_i(x)| \ge C-1] \le 2(\delta + \Pr[N \ge C-1])
$$
where $N \sim N(0,1)$ is a standard normal variable. Using
standard normal estimations
%(REF:http://godplaysdice.blogspot.com/2008/06/tail-bound-for-normal-distribution.html)
and setting $C = O(\sqrt{\log(m/\eps)})$ gives
$$
\Pr[|g_i(x)| \ge C-1] \le 2(\delta + \eps/100m)
$$
since $\delta < \eps/100m$ we get that $\Pr[|g_i(x)| \ge C-1] \le
\eps/10m$, and using the union bound over all $g_1,\ldots,g_m$ we
get that the total error is bounded by $\eps/10$.
\end{proof}

The following lemma bounds the tail moments of linear functions, and is somewhat similar to Lemma 4.2 in~\cite{DGJSV09}.

\begin{lem}\label{lem:bound_moments_tail}
Let $g:\{-1,1\}^n \to \R$ be a linear function with $\E[g^2]=1$.
Let $c > 0$ and $A \ge 2c$. Then
$$
\E_{x \in \{-1,1\}^n}[|g(x)|^{cA} 1_{|g(x)| \ge A}] \le 3 e^{2cA \ln(A) + 2c^2-\tfrac{1}{2} (A-2c)^2}.
$$
\end{lem}

\begin{proof}
Define $E_t = \E_{x \in \{-1,1\}^n}[|g(x)|^{cA} 1_{i \le |g(x)| <i+1}$. We have to bound
$E = \sum_{i \ge A} E_i$. By Hoeffding bound (see, e.g.,~\cite{AS08}),
$$
\Pr_{x \in U}[|g(x)| \ge i] \le 2 e^{-i^2/2}.
$$
Hence we get $E_i \le 2 e^{-i^2/2} (i+1)^{cA}$. Therefore
$$
(i+1)^{cA} \le i^{2cA} = A^{2cA} \cdot (i/A)^{2cA} \le A^{2cA} \cdot e^{2ci}
$$
where we used the fact that $x \le e^x$ for $x=i/A$. Summing over $i \ge A$ we get
\begin{align*}
E \le & A^{2cA} \sum_{i \ge A} e^{-i^2/2 + 2ci} = \\
&A^{2cA} \sum_{i \ge A} e^{-\tfrac{1}{2}(i-2c)^2 + 2c^2} \le \\
&3 A^{2cA} e^{2c^2} e^{-\tfrac{1}{2}(A-2c)^2}).
\end{align*}
where we used the fact that $\sum_{i \ge C} e^{-\tfrac{1}{2}i^2} \le \sum_{i \ge C^2} e^{-\tfrac{1}{2}i}
\le 3 e^{-C^2/2}$.
\end{proof}

We are now ready to prove Lemma~\ref{lem:approx_thresh}.

\begin{proof}[Proof of Lemma~\ref{lem:approx_thresh}]
Set $A>1$ to be determined later. Let $\tilde{G}:\R^m \to [-1,1]$
be the Lipschitz function approximating and bounding $\sgn(G)$
guaranteed by Lemma~\ref{lem:lip_approx}. Let $p:\R^m \to \R$ be
the polynomial guaranteed by Lemma~\ref{lem:approx_and_bound}
approximating $\tilde{G}$ on $[-A,A]^m$ with error $\eps/10$. The
degree of $p$ is $k_1 = O(\frac{A m^{3/2} L(\tilde{G})}{\eps}) = A
\cdot \phi(\eps)$, where $\phi(\eps) = O(\frac{m^{5/2} L}{\alpha
\eps})$ is independent of our choice of $A$. Set $p^*:\{-1,1\}^n
\to \R$ to be defined as
$$
p^*(x) = p(g_1(x),\ldots,g_m(x)).
$$
We have that
\begin{itemize}
    \item The polynomial $p^*$ is of degree at most $A \cdot
    \phi(\eps)$.
    \item For all $x \in \{-1,1\}^n$, $p^*(x) \ge \sgn(f(x))$.
    \item For all $x \in [-A,A]^m$ such that
    $(g_1(x),\ldots,g_m(x)) \notin \C(\mathcal{Z}(G),\tau)$ we have
    $p^*(x) \le \sgn(f(x)) + \eps/10$.
    \item For all $x \in [-A,A]^m$ such that
    $(g_1(x),\ldots,g_m(x)) \in \C(\mathcal{Z}(G),\tau)$ we have
    $p^*(x) \le 2$.
\end{itemize}

To conclude the proof we have to show that $\E_{x}[p^*(x) - \sgn(f(x))] \le \eps$. We consider three ranges of values for $x$.
\begin{enumerate}
    \item $x \in \{-1,1\}^n$ such that
    $(g_1(x),\ldots,g_m(x)) \in [-A,A]^m \setminus
    \C(\mathcal{Z}(G),\tau)$. \label{item1-approx}
    \item $x \in \{-1,1\}^n$ such that
    $(g_1(x),\ldots,g_m(x)) \in [-A,A]^m \bigcap
    \C(\mathcal{Z}(G),\tau)$.  \label{item2-approx}
    \item $x \in \{-1,1\}^n$ such that
    $(g_1(x),\ldots,g_m(x)) \notin [-A,A]^m$.  \label{item3-approx}
\end{enumerate}

To bound~\ref{item1-approx}, we use the fact that for
all $x$ such that $(g_1(x),\ldots,g_m(x)) \in [-A,A]^m \setminus
\C(\mathcal{Z}(G),\tau)$ we know that $p^*(x) - \sgn(f(x)) \le
\eps/10$, hence the total contributed error is bounded by
$\eps/10$.

To bound~\ref{item2-approx}, we use
Lemma~\ref{lem:zeros_nbd_small_prob} to conclude that the
probability over $x \in \{-1,1\}^n$ that $(g_1(x),\ldots,g_m(x))
\in \C(\mathcal{Z}(G),\tau)$ is bounded by $\eps/10$. Since we
know that for such $x$ we have $p^*(x) \le 2$ and $\sgn(f(x)) \ge
-1$, we can bound the total error by $3/10 \eps$.

Finally, let $\eps_3$ be the error in~\ref{item3-approx}. Namely,
$$
\eps_3 = \E_{x}\left[\left( p(g_1(x),\ldots,g_m(x)) - \sgn(f(x))
\right) \cdot 1_{(g_1(x),\ldots,g_m(x)) \notin [-A,A]^m}\right].
$$

We bound $\eps_3$ by the union bound over which of
$g_1(x),\ldots,g_m(x)$ is maximal.

$$
\eps_3 \le \sum_{i=1}^m \E_{x} \left[\left(
p(g_1(x),\ldots,g_m(x)) - \sgn(f(x)) \right) \cdot
1_{(g_1(x),\ldots,g_m(x)) \notin [-A,A]^m} \cdot 1_{g_i(x) =
\max(g_1(x),\ldots,g_m(x))} \right].
$$
Since $|p(z)| \le 2$ for $z \in [-1,1]^m$ and $|\sgn(f(x))|=1$,
by Lemma~\ref{lem:dubiner_md} we get
\begin{align*}
\eps_3 & \le  \sum_{i=1}^m \E_{x} \left[ (2 |2 g_i(x)|^{\deg(p)} +
1) \cdot 1_{|g_i(x)| \ge A} \right]\\
& \le 2^{\deg(p)+2} \sum_{i=1}^m \E_{x} \left[ |g_i(x)|^{\deg(p)}
 \cdot 1_{|g_i(x)| \ge A} \right]
\end{align*}

Recall that $\deg(p)=k_1=A \cdot \phi(\eps)$. Using
Lemma~\ref{lem:bound_moments_tail} we get the bound
$$
\eps_3 \le 3m 2^{cA} e^{2c A \ln{A} + 2c^2 - \tfrac{1}{2}
(A-2c)^2}
$$
where $c=\phi(\eps)$. Recall that $\phi(\eps) > m/\eps$, hence we
get that picking $A = \Omega(c \ln{c}) = \Omega(\phi(\eps)
\ln(\phi(\eps)))$ will yield $\eps_3 \le \eps/10$.
\end{proof}

\paragraph{Acknowledgement.} We are grateful to Moshe Dubiner for his great help with 
approximation theory and in particular in proving Lemma~\ref{lem:jackson_global_bound}.

\end{document}